\documentclass[pre,nofootinbib,floatfix,superscriptaddress, twocolumn]{revtex4}
\usepackage{dcolumn}
\usepackage{graphicx}
\usepackage{amsmath,amsfonts,amssymb, epsfig}
\usepackage{amsthm}
\usepackage{natbib}
\usepackage{epstopdf}
\usepackage{verbatim}
\usepackage[caption=false]{subfig}
\usepackage{color}
\usepackage{rotating}

\usepackage[pdftex,bookmarks, colorlinks, citecolor =blue, breaklinks]{hyperref}

\newcommand{\twopartdef}[8]
{
	\left\{
		\begin{array}{ll}
			#1 & \mbox{#2 } #3 \enspace#7 \\
			#4 & \mbox{#5 } #6 \enspace#8
		\end{array}
	\right.
}

\begin{document}
\title{Metabolite transport through glial networks stabilizes the dynamics of learning}
\author{Yogesh S. Virkar}
\email{Yogesh.Virkar@colorado.edu}
\affiliation{University of Colorado at Boulder, Boulder, CO, 80309, USA}
\author{Woodrow L. Shew}
\affiliation{University of Arkansas, Fayetteville, AR 72701, USA}
\author{Juan G. Restrepo}
\email{Juan.Restrepo@colorado.edu}
\affiliation{University of Colorado at Boulder, Boulder, CO 80309-0526, USA}
\author{Edward Ott}
\affiliation{University of Maryland, College Park, MD 20742, USA}

\begin{abstract}
Learning and memory are acquired through long-lasting changes in synapses \cite{feldman:2012}. In the simplest models, such synaptic potentiation typically leads to runaway excitation, but in reality there must exist processes that robustly preserve overall stability of the neural system dynamics. How is this accomplished? Various approaches to this basic question have been considered \cite{abbott:nelson:2000}. Here we propose a particularly compelling and natural mechanism for preserving stability of learning neural systems. This mechanism is based on the global processes by which metabolic resources are distributed to the neurons by glial cells. Specifically, we introduce and study a model comprised of two interacting networks: a model neural network interconnected by synapses which undergo spike-timing dependent plasticity (STDP); and a model glial network interconnected by gap junctions which diffusively transport metabolic resources among the glia and, ultimately, to neural synapses where they are consumed. Our main result is that the biophysical constraints imposed by diffusive transport of metabolic resources through the glial network can prevent runaway growth of synaptic strength, both during ongoing activity and during learning. Our findings suggest a previously unappreciated role for glial transport of metabolites in the feedback control stabilization of neural network dynamics during learning.
\end{abstract}


\maketitle

\section{Introduction}

\begin{figure}
\centering
\includegraphics[scale=0.9]{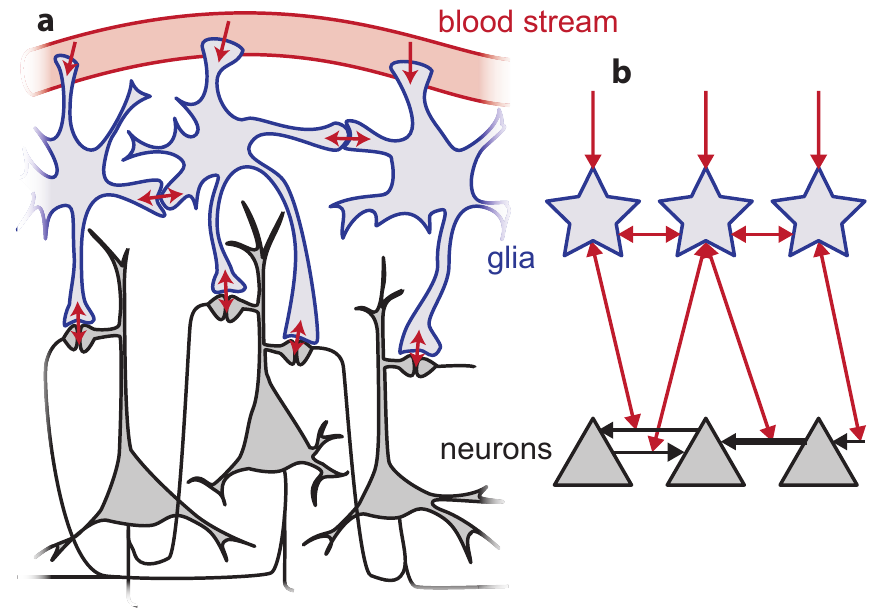}
\caption{\textit{Glial-neuronal interactions}: a) Cartoon based on existing experiments, illustrating how glia serve to distribute metabolic resources from the bloodstream to neural synapses. Red arrows indicate paths of metabolite transport.  b) A simplified directed graph representation of our two-layer network model.  Black arrows indicate neural synaptic interactions. Arrow thickness indicates synaptic strength which evolves according to STDP. Red arrows which terminate on black arrows represent the resource supply to the corresponding synapse.}
\label{fig:resource_glia_neuron}
\end{figure}

Glial brain cells play important and diverse roles regulating the dynamics and structure of neural networks \cite{giaume:et:al:2010, fields:woo:basser:2015}, including learning-related changes in synapses \cite{suzuki:et:al:2011, newman:korol:gold:2011}. In this paper we focus on one of the most important functions thought to be served by the glial network -- the transport and distribution of metabolic resources among the neural synapses \cite{rouach:et:al:2008}. This hypothesis originated from early anatomical studies which showed that the glia form a bridge between the neural synapses and the brain vasculature \cite{tsacopoulos:magistretti:1996} (Fig~\ref{fig:resource_glia_neuron}a). More recently, experiments have directly demonstrated that glia, astrocytes more specifically, deliver metabolic resources to synapses depending on how active the synapses are \cite{pellerin:magistretti:1994}. En route to the synapses, these resources diffuse through an extensive network of astrocytes \cite{rouach:et:al:2008}.  The biophysical properties of such diffusive transport of resources may have a fundamental influence on the dynamics of the activity of the neural network that consumes the resources \cite{wade:et:al:2014, gordleeva:et:al:2012, jolivent:et:al:2015, simpson:carruthers:vannucci:2007}.  For example, a highly active synapse may consume all of its local resources, thus forcing it to become less active until more resources arrive, and may drain resources away from less active synapses, thus shaping functional differences among synapses.  Here, in order to study these possibilities, we introduce a computational model incorporating both a neural network and a glial network.  
Our model neurons interact via synapses whose efficacy evolves according to activity-dependent learning rules, namely spike timing dependent plasticity (STDP) \cite{feldman:2012, abbott:nelson:2000}.  Under many circumstances, modeling of STDP can result in unstable growth of synaptic efficacy and typically requires additional types of learning rules to prevent such runaway growth (see discussion in the Conclusion section and Refs.~\cite{abbott:nelson:2000, delattre:et:al:2015}).  The main finding of our work is that diffusive transport of resources via the glial network can serve to prevent runaway synaptic growth due to STDP, thereby maintaining stable neural network dynamics.  We show that this phenomenon requires resource transport among the glia; locally confined production and consumption of resources result in unstable neural network dynamics.  The known roles played by the glia in synaptic plasticity are diverse and numerous \cite{min:santello:nevian:2012}, but, to our knowledge, our work is the first to show that metabolic resource distribution can play such a stabilizing role. 

More broadly, it is interesting to note the very general finding that many complex systems arising in diverse fields (e.g., social interactions, biology, information science, etc.) are founded upon interactions between two or more identifiable networks. For a recent review with useful further references and examples see Ref.~\cite{kivela:2014}. Our work in the present paper provides a compelling example of the potential dynamical benefit of this type of structure in the architecture of the brain.

\section {Model}
As shown in Fig \ref{fig:resource_glia_neuron}b, our model consists of a two-layered network whose first layer is a weighted and directed neural network and whose second layer is an unweighted undirected glial network. 

The neural network is composed of $N$ excitable nodes that represent neurons, labeled $n = 1,2,\dots, N$, and $M$ directed edges, labeled $\eta = 1,2, \dots, M$ on which synapses are located. The state $s_n^t$ of neuron $n$ at a discrete time step $t$ is represented either as $s_{n}^t = 0$ (quiescent) or $s_{n}^t = 1$ (active). We define $W^t$ as the $N \times N$ adjacency matrix whose entry $W_{nm}^t$ denotes the weight of the synapse on the edge from neuron $m$ to neuron $n$ at time $t$. Any presynaptic neuron $m$ can be either excitatory ($\epsilon_m = 1$) or inhibitory ($\epsilon_m = -1$). Thus, if we let $w_{nm}^t = |W_{nm}^t|$ denote the absolute value of synapse strength, then $W_{nm}^t = \epsilon_m w_{nm}^t$. 

At each time step $t$ (where $t=0, 1, 2, \dots$), the state of neuron $n$ is updated probabilistically based on the sum of its synaptic input from active presynaptic neurons in the previous time step, 
\begin{align} \label{eq:sn_t_1}
s_n^{t+1} = \twopartdef{1}{with probability}{\sigma \left(\displaystyle\sum_{m=1}^{N} W_{nm}^t s_m^t\right)} {0} {otherwise} {}{,}{.}
\end{align}
As in Ref.~\cite{larremore:et:al:2014}, the model transfer function probability $\sigma$ is piecewise linear; $\sigma(x) = 0$ for $x \leq 0$, $\sigma(x) = x$ for $0 < x < 1$, and $\sigma(x) = 1$ for $x \geq 1$. 

The second layer of our model, the unweighted and undirected glial network, consists of $T$ glial cells labeled $i = 1,2,\dots, T$. Each glial cell $i$ holds an amount of resource $R_{i}^t$ at time step $t$. While in this paper we do not focus on a particular resource, we note that various metabolites are transported diffusively among the glial cells such as glucose and lactate \cite{rouach:et:al:2008}. Resources diffuse between the glial cells that are connected to each other. We define a $T \times T$ symmetric glial adjacency matrix $U$ such that entry $U_{ij} = 1$ if glial cell $j$ is connected to glial cell $i$ and $U_{ij} = 0$ otherwise. Each glial cell serves a set of synapses by supplying resource to them. Hence we define a $T \times M$ matrix $G$ with entries $G_{i\eta} = 1$ if glial cell $i$ serves synapse $\eta$ and $G_{i \eta} = 0$ otherwise. Consistent with recent experimental studies \cite{halassa:et:al:2007}, we assume that each synapse is served by a unique glial cell and that all incoming synapses of one neuron (i.e., its dendrites) are served by one glial cell. So, given a synapse $\eta$, there is a unique glial cell $i(\eta)$ such that $G_{i(\eta)\eta}$ = $1$. 

{\textit{Learning}}: Let $\eta$ denote the synapse that connects presynaptic neuron $m$ to postsynaptic neuron $n$, i.e., the synapse $\eta$ that corresponds to the neural network edge $m \to n$. We assume that the absolute  strength of synapse $\eta$, i.e., $w_{nm}$, depends on its past learning history as determined from the STDP learning rule via an auxiliary variable, $\widehat{w}_{nm}^t$, and on the amount of resource $R_{\eta}^t$ at synapse $\eta$, 
\begin{align} \label{eq:wnm_learning_resource}
w_{nm}^t = f\left(R_{\eta}^t,\enspace{\widehat{w}}_{nm}^t\right) \enspace,
\end{align}
where $\partial f(x, y)/\partial x \geq 0$, $\partial f(x,y)/\partial y \geq 0$, and $\widehat{w}_{nm}^{t}$ evolves according to the STDP learning rule: 
\begin{align}\label{eq:stdp_ssl}
\widehat{w}_{nm}^{t+1} = \widehat{w}_{nm}^t \exp\left[{\frac{\epsilon_m}{\tau} \left(s_{m}^{t-1} s_{n}^{t} - s_{m}^{t} s_{n}^{t-1} \right)}\right] \enspace.
\end{align}
Moreover, we implement synaptic strength limitation, by requiring $f$ not to exceed a maximum value $\bar{w}$, $f \leq \bar{w}$. For excitatory synapses ($\epsilon_m = +1$), causal firing corresponds to firing of the presynaptic neuron $m$ on the previous time step $t-1$ (i.e., $s_{m}^{t-1} = 1$), followed by the firing of the postsynaptic neuron $n$ on the current time step $t$ (i.e., $s_{n}^t = 1$). Thus for causal excitations $\widehat{w}_{nm}^{t+1} > \widehat{w}_{nm}^{t}$ and the excitatory synapse is reinforced. Similarly, for anticausal excitations excitatory synapses are weakened, $\widehat{w}_{nm}^{t+1} < \widehat{w}_{nm}^{t}$. The corresponding analogous conditions hold for inhibitory neurons ($\epsilon_m = -1$). The constant $\tau$ sets the learning timescale. 

\begin{figure*}
\centering{
\subfloat{{\label{fig:stability_exp_lambda}}\includegraphics[scale=0.015]{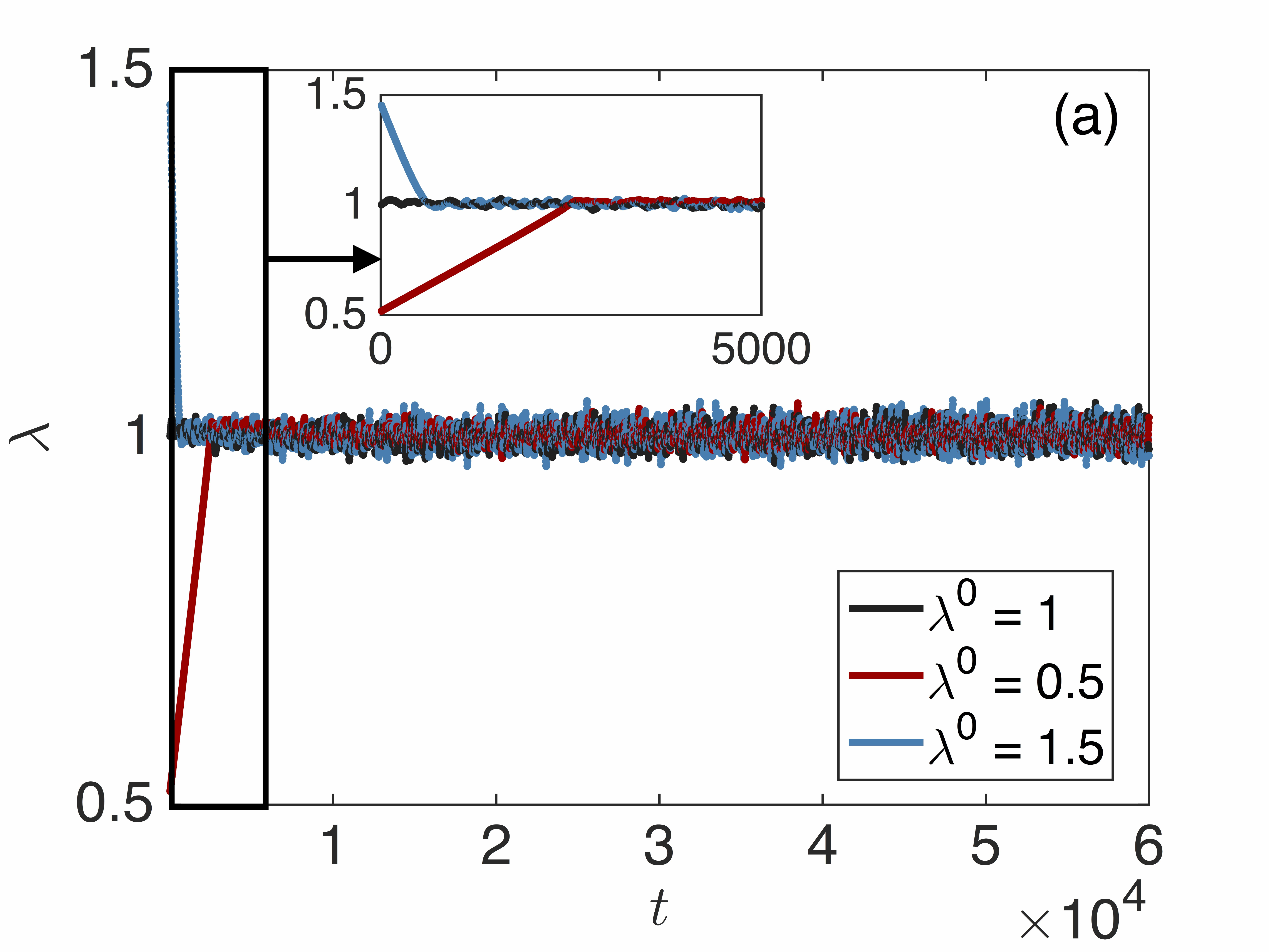}} 
\subfloat{{\label{fig:stability_exp_R}}\includegraphics[scale=0.31]{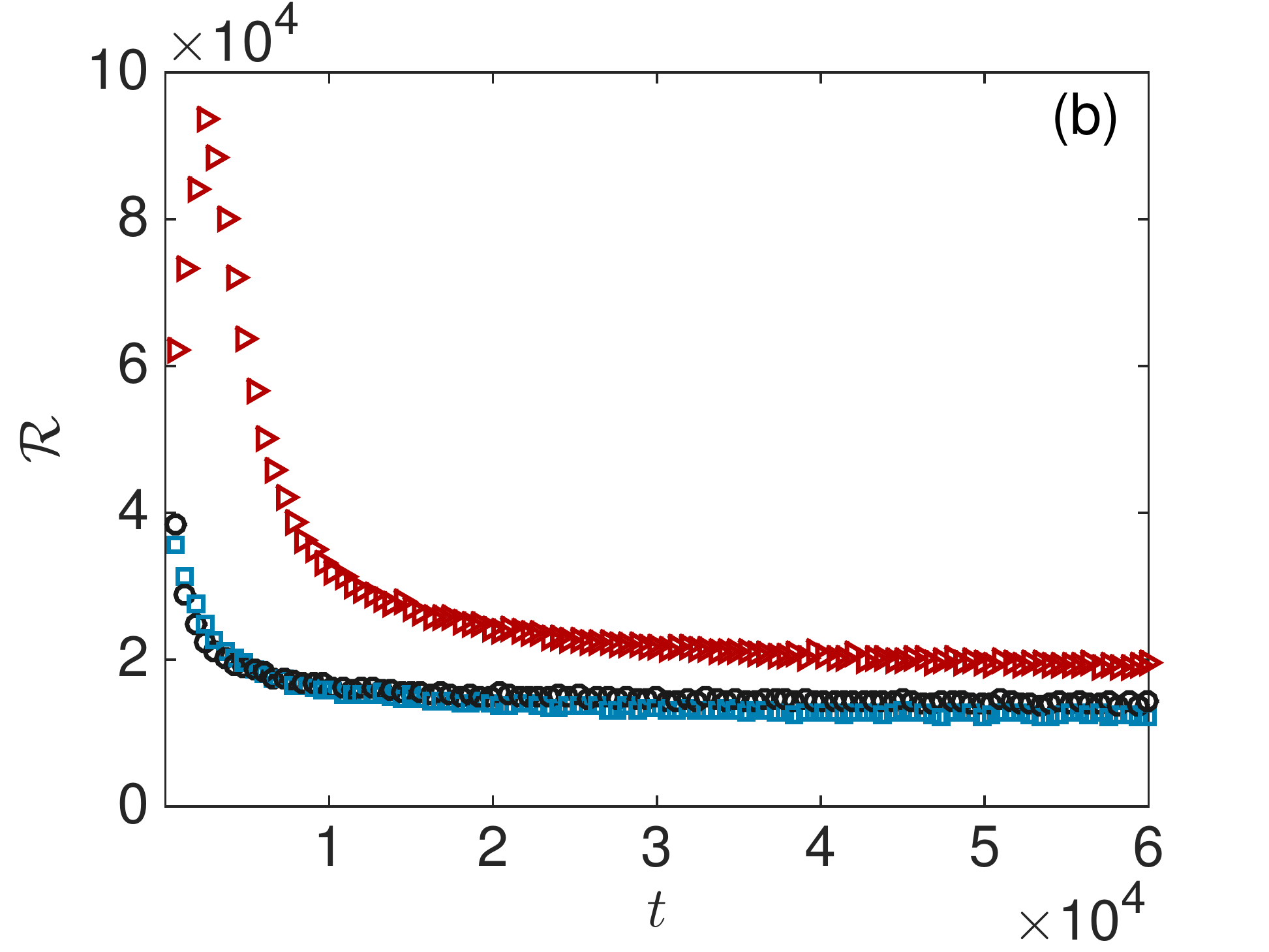}}
\subfloat{{\label{fig:stability_S}}\includegraphics[scale=0.31]{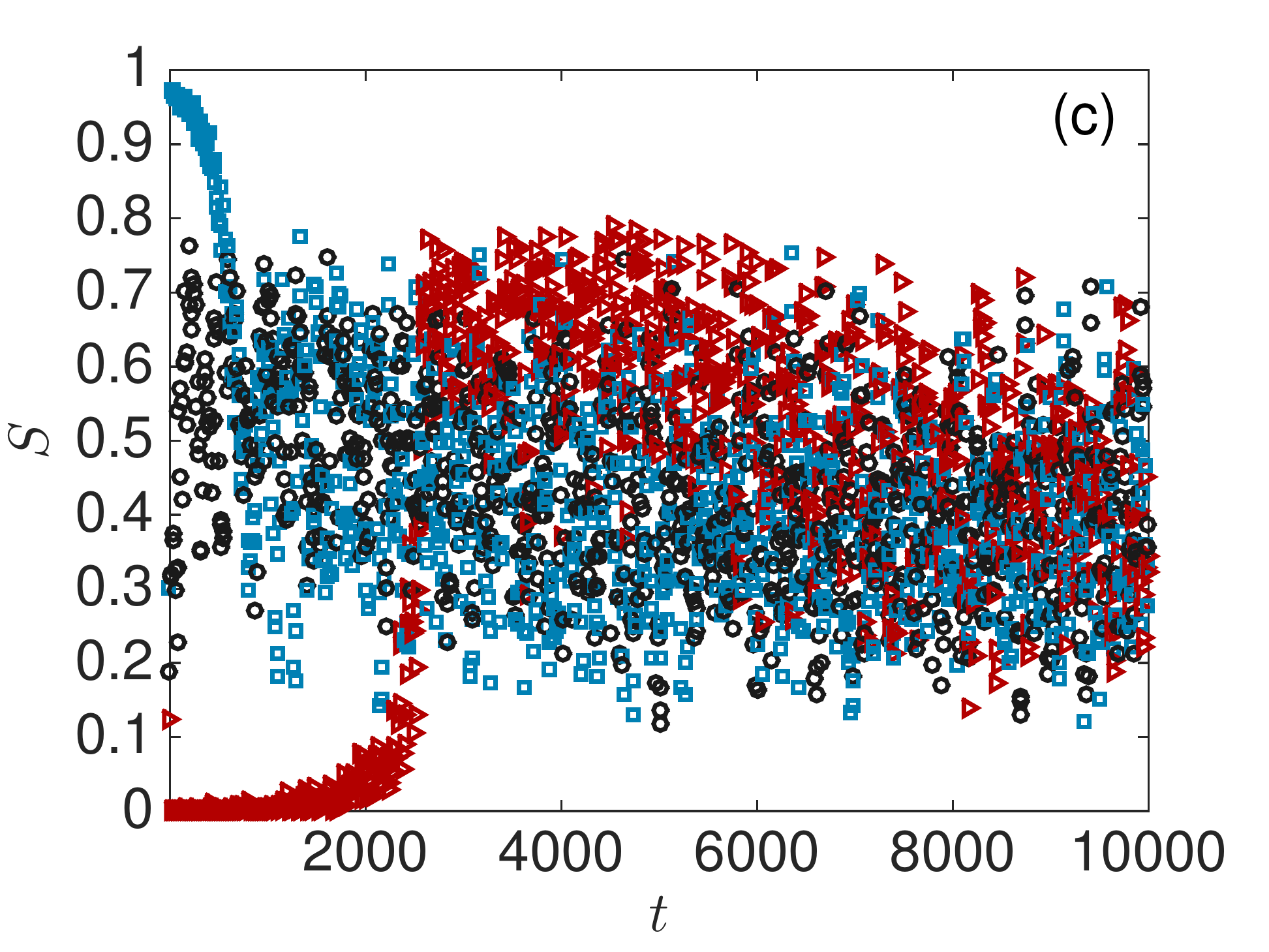}}
}
\caption{\textit{Resource-transport dynamics stabilizes network activity (Experiment 1)}: (a) Time series of $\lambda^t$ (largest eignenvalue of $W^t$) reveal rapid convergence to stable network dynamics ($\lambda \approx 1$), independent of initial conditions.  Three different initial conditions were tested: hyperexcitable (blue,  $\lambda^0 = 1.5$), stable (black,  $\lambda^0 = 1$), and hypoexcitable (red, $\lambda^0 = 0.5$). The inset is an expanded view of the first $5000$ time steps. (b) After a longer transient the total resource $\mathcal{R}$ also stabilizes to a steady value. (c) Similarly, in all three cases, the average activity $S$ reaches a statistical steady state with large fluctuations.} 
\label{fig:stability_exp}
\end{figure*}

{\textit{Resource-transport dynamics}}: Resource diffuses between glia through their connection network (characterized by the adjacency matrix $U$) and between glia and the synapses they serve (via the glial-neural connection network characterized by the adjacency matrix $G$). Our model for the evolution of the amount of resource $R_i^t$ at glial cell $i$ and the amount of resource $R_{\eta}^t$ at synapse $\eta$ is 
\begin{align}\label{eq:diffusion_betweenGlia}
R_i^{t+1} &= R_i^t + C_1 + D_G \displaystyle\sum_{j=1}^T U_{ij} \left(R_j^t - R_i^t\right) \nonumber \\ &+ D_S \displaystyle\sum_{\eta = 1}^{M} G_{i\eta} \left(R_{\eta}^t - R_{i}^t \right)  \enspace,
\end{align}
\begin{align}\label{eq:diffusion_betweenGliaSynapse}
R_{\eta}^{t+1} = R_{\eta}^t + D_S \left(R_{i(\eta)}^t - R_{\eta}^t \right) - C_2 s_{m(\eta)}^t \enspace,
\end{align}
where $D_G$ is the rate of diffusion between glial cells, and $D_S$ is the rate of diffusion between glia and synapses. Moreover, we enforce $R_{\eta} \geq 0$, i.e., if \eqref{eq:diffusion_betweenGliaSynapse} yields $R_{\eta}^{t+1} < 0$, then we replace it by $0$. The first term on the right hand side of \eqref{eq:diffusion_betweenGlia}, $R_i^t$, is the amount of resource in glial cell $i$ at time $t$. The parameter $C_1$ denotes the amount of resource added to each glial cell at each time step  (e.g., supplied by capillary blood vessels). For simplicity, we assume each glial cell has the same $C_1$. The last two terms are the amount of resource transported to glial cell $i$, respectively, from its neighboring glial cells and from the synapses that it serves.

In \eqref{eq:diffusion_betweenGliaSynapse}, the first term denotes the amount of resource at synapse $\eta$ at time $t$. The term proportional to $D_S$ denotes the amount of resource gained (if $R_{i(\eta)}^t > R_{\eta}^t$) or lost (if $R_{i(\eta)}^t < R_{\eta}^t$) from glial cell $i(\eta)$ that serves synapse $\eta$. If the presynaptic neuron $m(\eta)$ fires at time step $t$ ($s_{m(\eta)}^t = 1$), then all outgoing synapses for neuron $m(\eta)$, including $\eta$, consume some resource, thus decreasing the resource at each synapse by an amount $C_2$ (where $C_2$ is a model parameter).

\section{Numerical experiments and results}

In this section, we present results of numerical experiments on our model. For simplicity we assume that both the neural network and the glial network have an Erd\"os-Renyi (ER) network structure. We build the $N\times N$ directed weighted ER neural network adjacency matrix $W$ by creating a link from node $m$ to node $n$ (i.e., setting $W_{nm} \neq 0$) with probability $p$ and setting $W_{nm} = 0$ otherwise. This gives the mean number of incoming and outgoing synapses per neuron, $k_N = Np$, and the expected total number of synapses $M = N k_N $. To specify the initial state of each synapse, at $t=0$ we set $R_{\eta}^0 = 1$ and take the initial value of each $\widehat{w}_{nm}^0$ to be an independent draw from a uniform distribution over $[0, 1]$. We then rescale all entries in $W$ by a constant to obtain a desired largest eigenvalue of $W$, as discussed below. 

The glial network, represented by the matrix $U$ having $T$ nodes that represent glial cells, is taken to be an undirected and unweighted ER network. If the glial cell $j$ is connected to the glial cell $i$, then $U_{ij} = U_{ji} = 1$; and $U_{ij} = 0$ otherwise. If the probability of forming a link is $q$, then the mean degree of a glial cell is $k_G = Tq$. Recent evidence suggests that the number of glial cells are roughly equal to the number of neurons \cite{azevedo:et:al:2009}, and hence in our experiments we set $T = N$. The initial resource for each glial cell is taken to be $R_i^0  = 1$. 

In all our experiments we take the function $f$ in \eqref{eq:wnm_learning_resource} to be $f(x,y) = xy$ for $xy < \bar{w}$ and $f(x,y) = \bar{w}$ for $xy \geq \bar{w}$. We set $N=1000$ and $p=0.05$ and randomly draw an ER directed random graph for the neural network. We make another draw for the undirected glial network with $T = N = 1000$ and $q = p = 0.05$. This gives us $k_N = k_G = 50$. For all our numerical experiments we take $D_G$ and $D_S$ to be the same, $D_G=D_S=D$; we also set the fraction of inhibitory nodes to be $0.2$ \cite{meinecke:peters:1987} and use the following additional parameter choices
\begin{align}\label{eq:parameterChoices}
C_1 &= 0.0188, \enspace C_2 = 0.001, \nonumber\\
D &= 0.005 \enspace, \bar{w} = 0.14. \nonumber
\end{align}
We chose these parameter values somewhat arbitrarily but, as shown later, our results are fairly robust to the choice of these values. 

In the following, we report the three main findings from our model.  First, we show that network dynamics are stable, avoiding saturation or extinction of neural activity.  Second, we show that resource transport among the glia is essential to maintain this stability.  Third, we verify that the neural network can learn, i.e., external input results in long-lasting synaptic changes.

\begin{figure*}
\centering{
\subfloat{{\label{fig:instability_exp_lambda}}\includegraphics[scale=0.31]{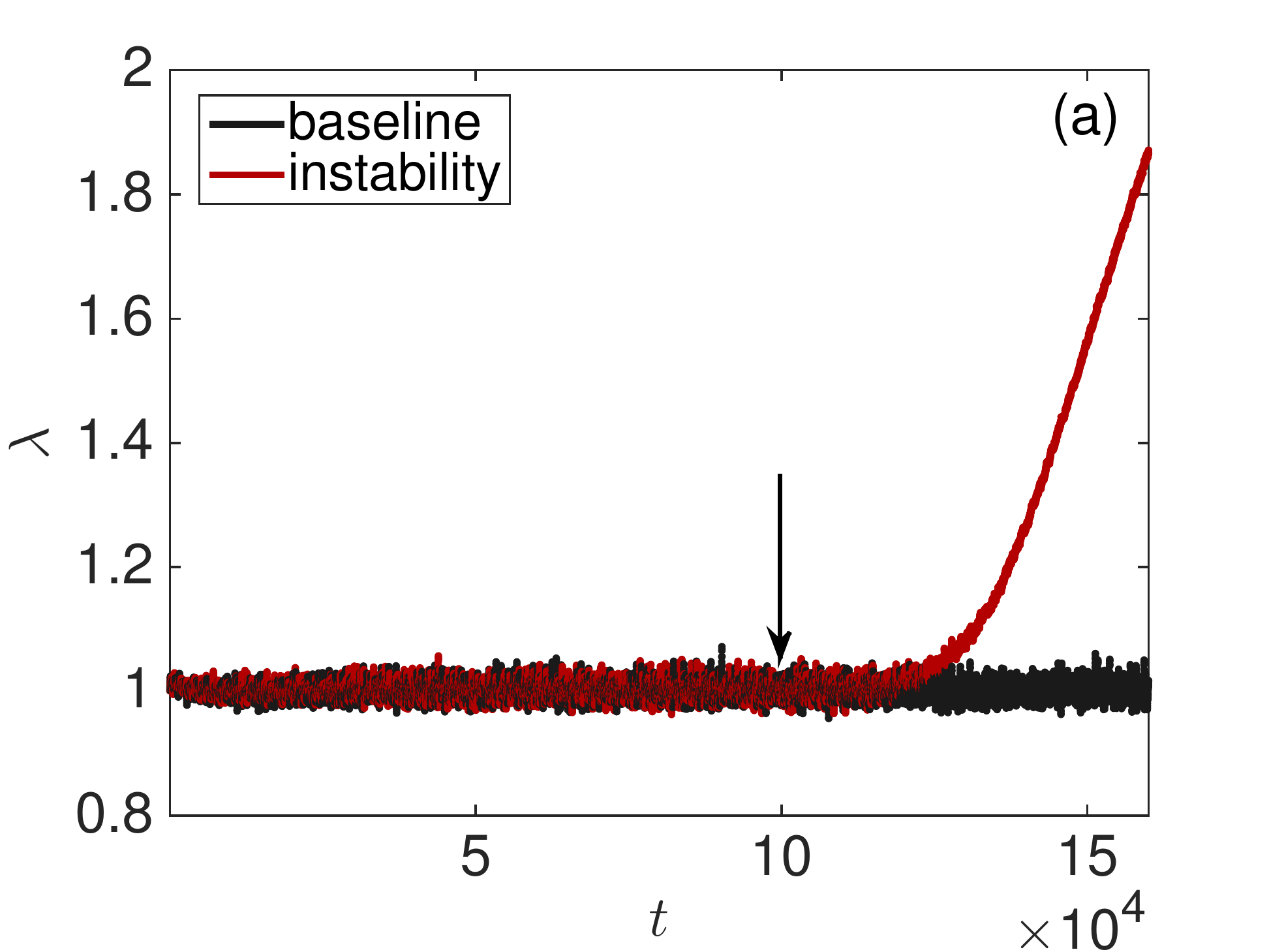}}
\subfloat{{\label{fig:instability_exp_R}}\includegraphics[scale=0.31]{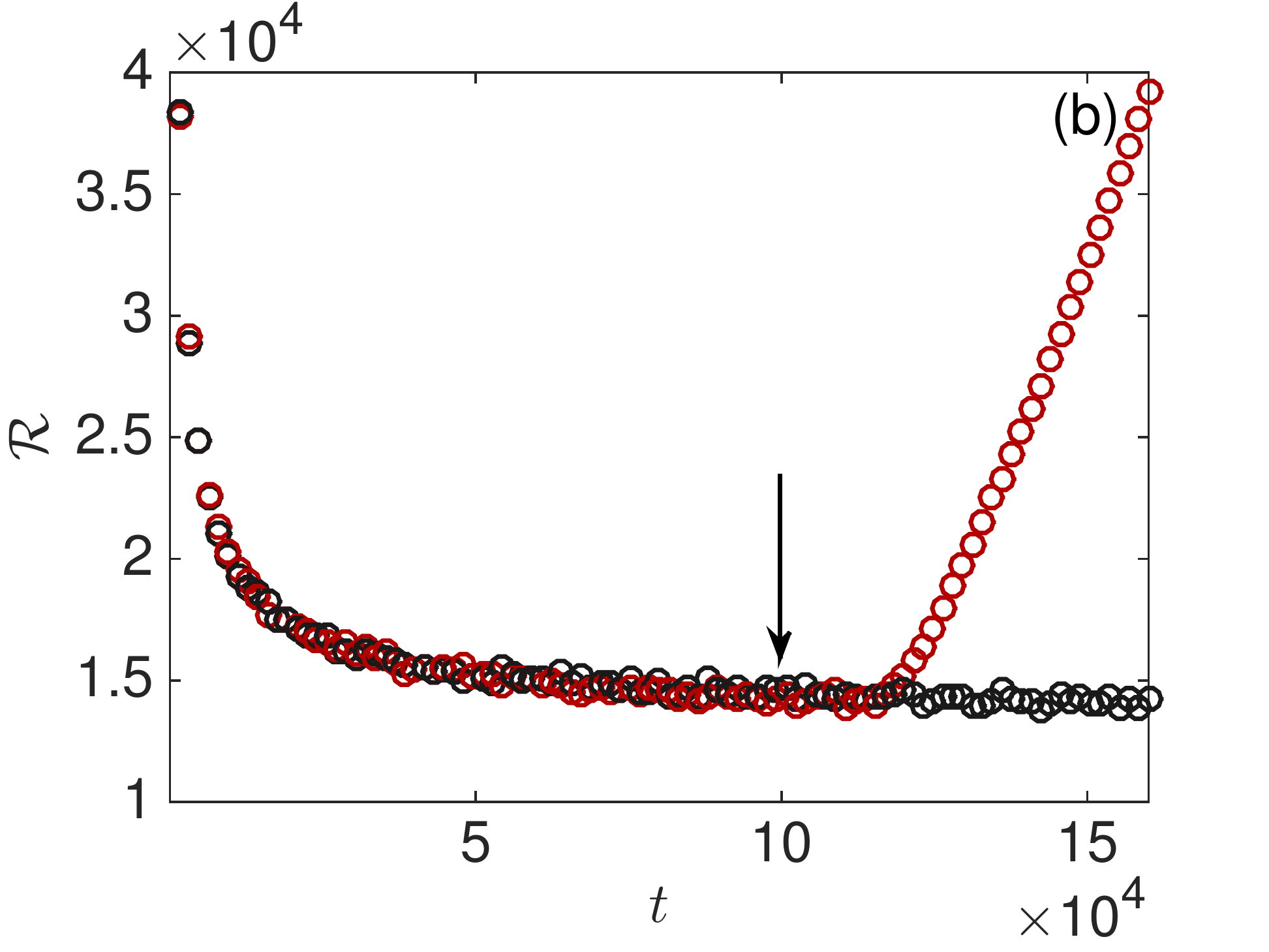}}
}
\caption{\textit{Turning off diffusion results in runaway growth (Experiment 2)}: (a) The maximum eigenvalue $\lambda$ versus $t$, and (b) the total resource $\mathcal{R}$ versus $t$. The data plotted in black are `baseline' results obtained using our model as described in the \textbf{Model} section of the paper. For the data plotted in red (labelled `instability'), the initial evolution is the same as for the baseline data up until $t=100000$ (marked in the figure by a vertical arrow), at which time the diffusion of resources between the glial cells is turned off, as described in the text. 
}
\label{fig:instability_exp}
\end{figure*}

\textit{Experiment 1}: To quantitatively assess the stability of the network dynamics we study $\lambda$, the largest eigenvalue of the matrix $W$.  Previous studies on purely excitatory networks \cite{larremore:shew:restrepo:2011} and networks having inhibitory nodes \cite{pei:et:al:2012, larremore:et:al:2014} show that $\lambda$ determines the nature of the network's dynamics: $\lambda< 1$ corresponds to a hypoexcitable, or subcritical, state where activity dies out; $\lambda = 1$ corresponds to the stable, critical state where activity is balanced, neither growing nor decaying on average; and $\lambda > 1$ corresponds to a hyperexcitable, supercritical state where the activity grows until nearly all neurons are firing at every time step. Moreover, it has been shown experimentally \cite{shew:plenz:2013} that criticality ($\lambda$ near $1$) provides neural networks with potential benefits for information processing. 
In this first experiment we choose different initial conditions for $\lambda$ (obtained by rescaling the initial $W$), i.e., at $t=0$ we start in the critical, subcritical and supercritical states respectively, $\lambda^0 = \{1, 0.5, 1.5\}$. Figure~\ref{fig:stability_exp_lambda} shows a plot of $\lambda$ as a function of time, $t$. In all three cases we find that after a brief transient, the network dynamics become stable, i.e., $\lambda$ fluctuates near 1 after sufficient time has passed. Also, starting at the critical state does not result in any instabilities over time. Fig.~\ref{fig:stability_exp_R} shows the total resource $\mathcal{R}$ held in all glia and synapses as a function of time $t$, where $\mathcal{R}$ is given by 
\begin{align}
\mathcal{R}^t = \sum_{i=1}^{T} R_i^t + \sum_{\eta=1}^{M} R_{\eta}^t\enspace.
\end{align}
In all three cases $\mathcal{R}$ reaches a steady state value. Fig.~\ref{fig:stability_S} shows that the average activity,   
\begin{align}
S = \displaystyle\frac{1}{N} \displaystyle\sum_{n=1}^{N} s_n^t\enspace,
\end{align}
is initially below the activity for the critical case  for $\lambda^0=0.5$ and above the activity for the critical case for $\lambda^0=1.5$, indicative of the subcritical and supercritical regimes. Starting in these regimes, over time, the dynamics of $S$ becomes statistically similar to the dynamics of $S$ for the critical initial state of $\lambda^0=1$. Thus our model naturally leads to a network that operates in a stable critical ($\lambda$ near $1$) regime. This can be understood on the basis that high activity rapidly consumes resources at the synapses, thus reducing their weights, and leading to decrease in $\lambda$;  while with low activity, synapses consume resource at a low rate, allowing buildup of resource with time and consequent increase of synaptic weights, essentially a feedback control stabilization process. 
An indication of the potential information handling benefits of criticality \cite{shew:plenz:2013, kinouchi:copelli:2006} can be seen in Fig.~\ref{fig:stability_S} at early time, $t \lesssim 300$, where we observe that in both the subcritical (red triangles) and supercritical (blue squares) states there is relatively little time variation, corresponding to relatively little potential for information content; while, in contrast, in the critical case (black circles), the signal varies over a larger range ($0.3 \lesssim S \lesssim 0.7$). 

\begin{figure*}
\centering{
\subfloat{{\label{fig:learning_protocol}}\includegraphics[scale=0.31]{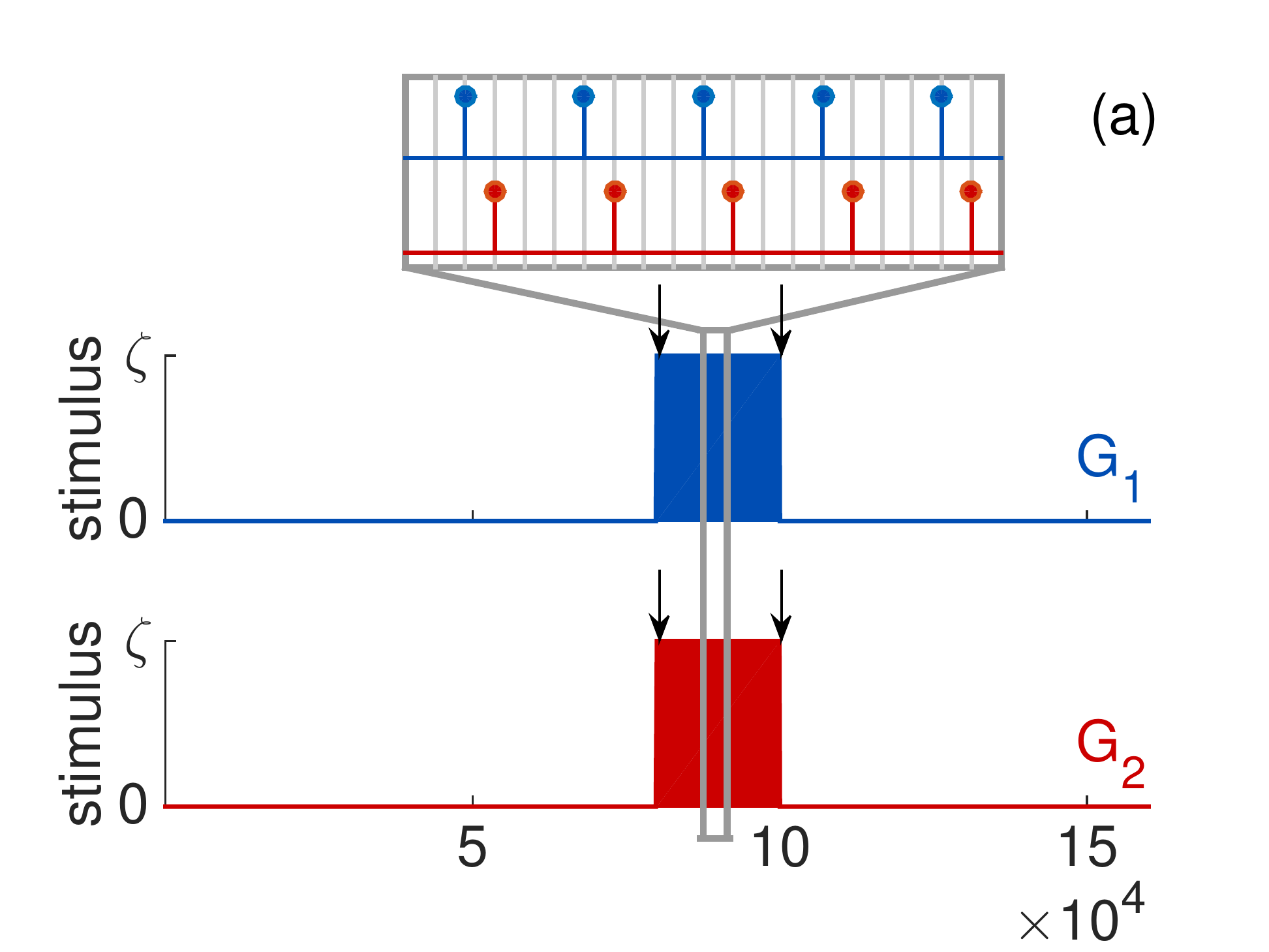}}
\subfloat{{\label{fig:learning_lambda}}\includegraphics[scale=0.31]{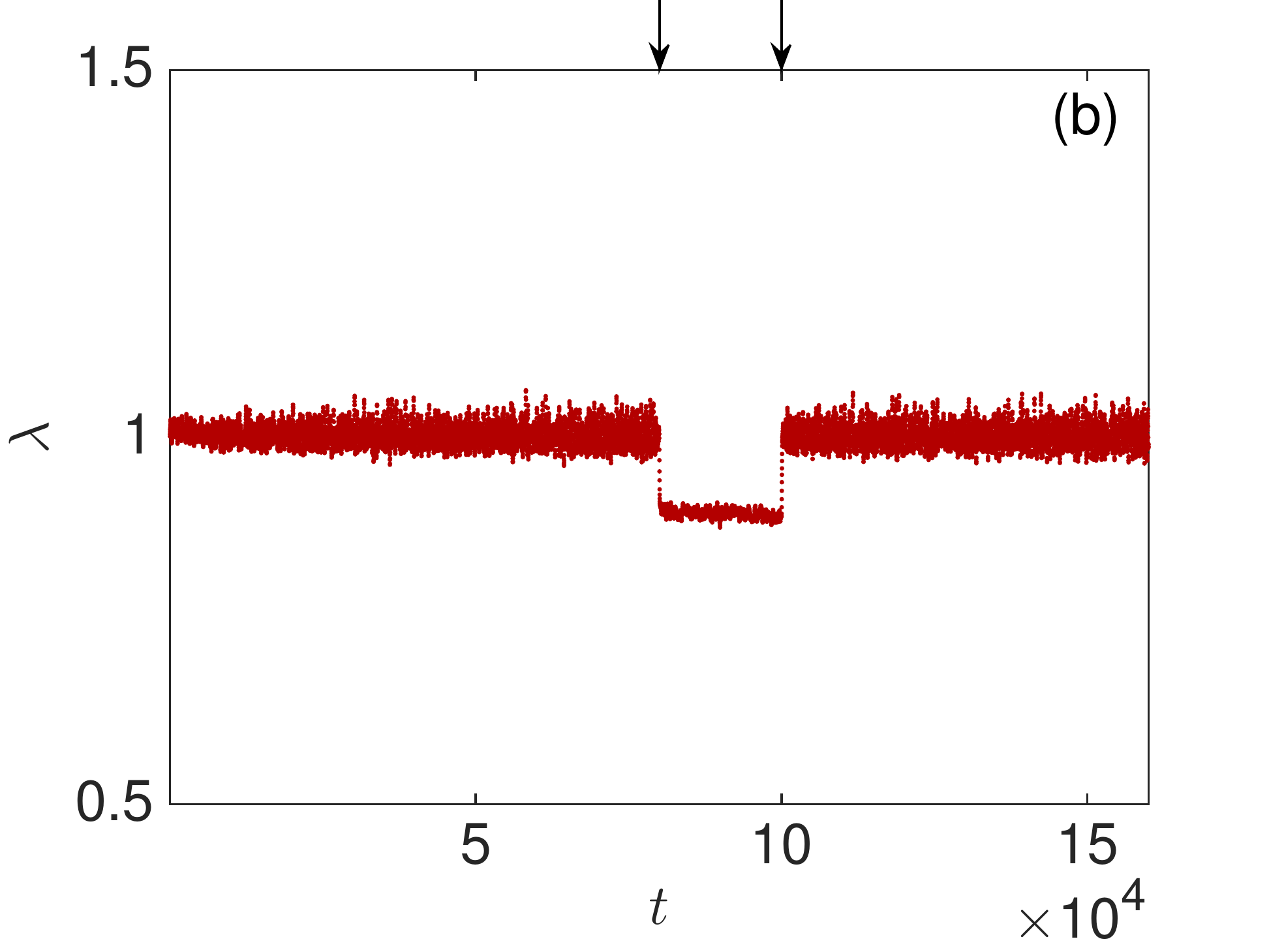}}\\
\subfloat{{\label{fig:learning_R}}\includegraphics[scale=0.31]{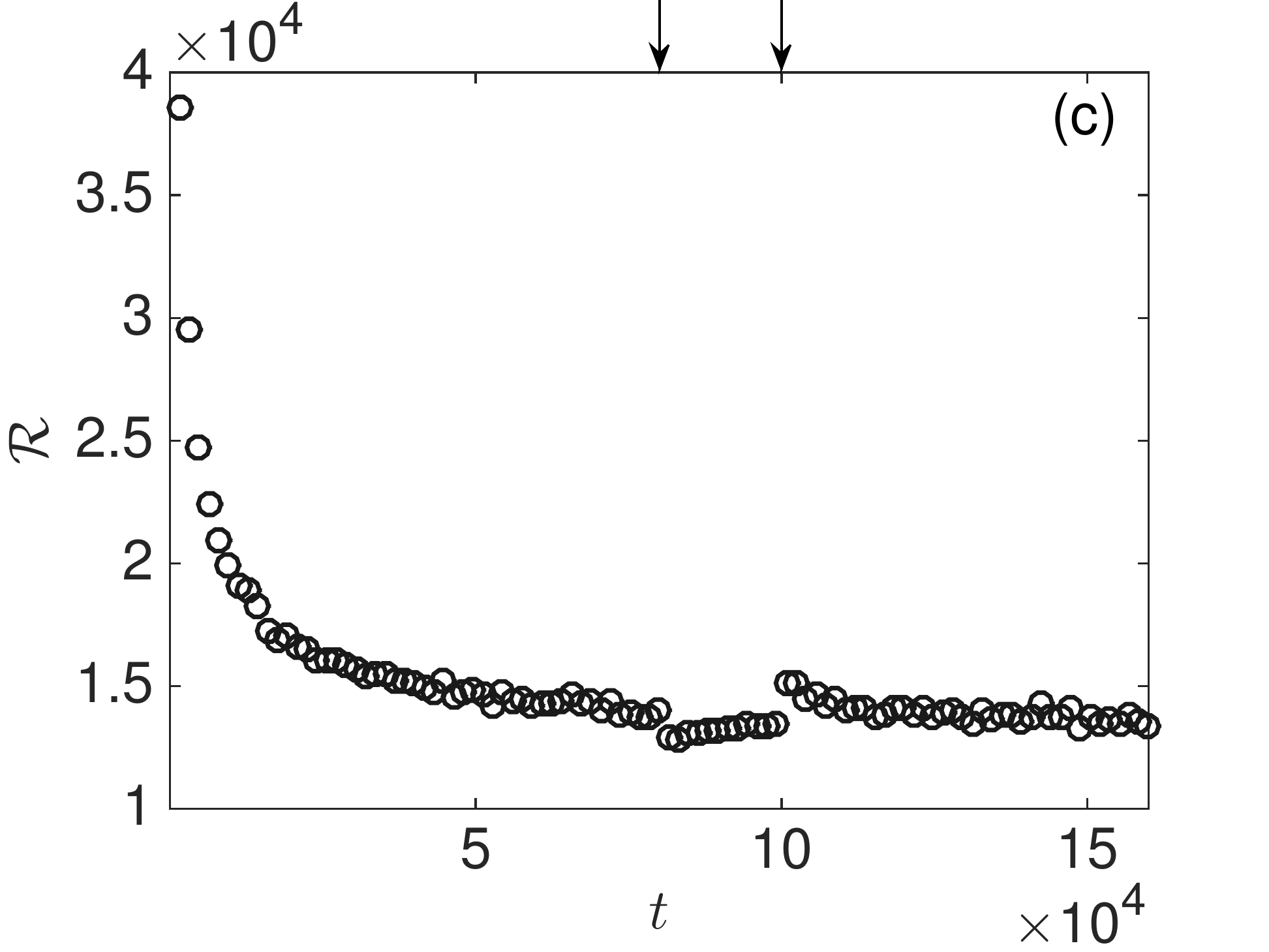}}
\subfloat{{\label{fig:learning_excitatory}}\includegraphics[scale=0.31]{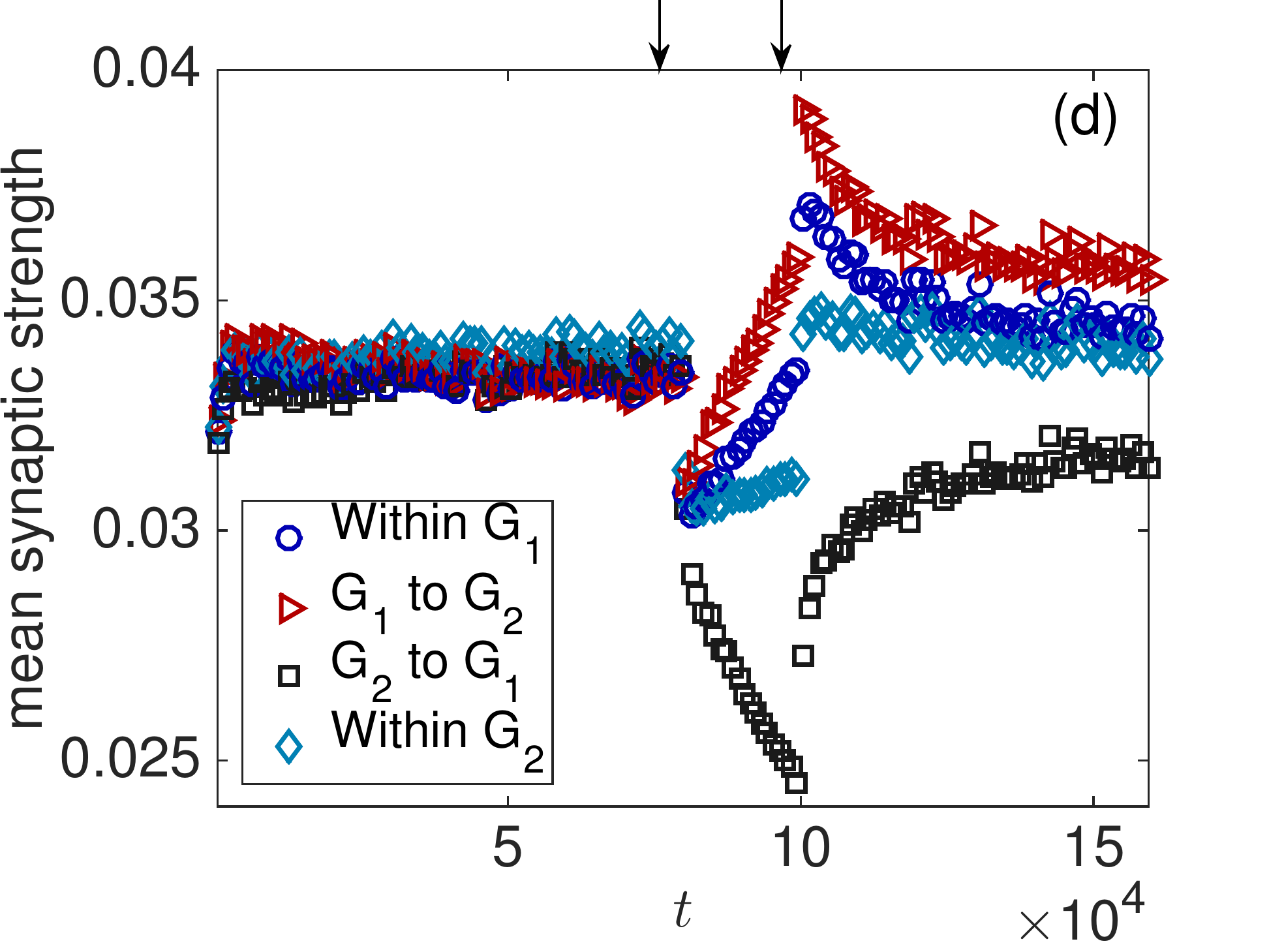}}
}
\caption{\textit{The STDP network learns and remembers (Experiment 3)}: We divide the neurons into two equally sized groups, $G_1$ and $G_2$, consisting of $500$ neurons each. This results in four groups of synapses: synapses within the first group (\textit{Within $G_1$}), synapses that convey signals from neurons in $G_1$ to neurons in $G_2$ (\textit{$G_1$ to $G_2$}), synapses that convey signals from neurons in $G_2$ to neurons in $G_1$ (\textit{$G_2$ to $G_1$}) and synapses within the second group (\textit{Within $G_2$}). Panel (a) depicts the learning protocol (see text). Panels (b) and (c) show $\lambda$ and $\mathcal{R}$ versus $t$. The learning regime spans $t=[80000, 100000]$ (delimited by the vertical arrows). Panel (b) shows that $\lambda$ becomes subcritical during learning \cite{fagerholm:et:al:2015}, but then quickly evolves back to the critical state $\lambda \cong 1$. Panel (d) shows the mean synaptic strength for the four groups of synapses for excitatory synapses during learning. In accord with the STDP learning rule, the mean synaptic strength increases for $G_1$ to $G_2$ synapses. In the post-learning regime, spanning $t=[80000, 160000]$, panel (d) shows that the model remembers what it learned. 
}
\label{fig:learning}
\end{figure*}

\textit{Experiment 2}:  STDP and resource distribution dynamics are both active during the stabilization demonstrated in Figure~\ref{fig:stability_exp_lambda}.  Next, we pose the question: Is the diffusion of resources via the glial network important for stable cortical dynamics? Or can we still get stability if we switch off transport among the glia (i.e., set $D_G = 0$)? To do this experiment, for $t=1, 2, \dots, T_1=80000$, we let the system reach a steady state with the glial network operative as in Fig.~\ref{fig:stability_exp}, using \eqref{eq:diffusion_betweenGlia}, and define an equivalent time averaged resource supply rate $C_i$ for each glial cell $i$, 
\begin{align} \label{eq:tunedSource}
C_i = \left\langle D_G \displaystyle\sum_{j=1}^T U_{ij} \left(R_j^t - R_i^t\right)  \right\rangle_{T_1, T_2} + C_1 \enspace . 
\end{align} 
In the above equation $\langle \rangle_{T_1, T_2}$ represents the time average over the interval $t = (T_1, T_2)$. We switch off transport among the glia at $t = T_2$ by setting $D_G = 0$, and replace \eqref{eq:diffusion_betweenGlia} by
\begin{align} \label{eq:renewedDiffusionEq}
R_i^{t+1} = R_i^t + D_S \displaystyle\sum_{\eta = 1}^{M} G_{i\eta} \left(R_{\eta}^t - R_{i}^t \right) + C_i \enspace.
\end{align}
Thus the average rate of total nonsynapse resource supply to each glial cell is preserved after the glial diffusion is turned off. Replacing \eqref{eq:diffusion_betweenGlia} by \eqref{eq:renewedDiffusionEq} after $t=T_2=100000$, we run the dynamics for a total of $160000$ time steps. 

For initial condition $\lambda^0 = 1$, Fig.~\ref{fig:instability_exp} shows the results for two runs-- one in which we use the dynamics described by Eq. (4) (baseline) and the other in which we switch off the glial network and run the dynamics as described above (instability). Fig.~\ref{fig:instability_exp_R} shows that after the glial network is switched off, $\mathcal{R}^t$ increases as resource starts to accumulate at some synapses and gets used up at others.  Such increases and decreases in $R_{\eta}$ change the weights of the matrix $W$ resulting in an increase in $\lambda$ as shown in Fig. \ref{fig:instability_exp_lambda}. Thus the dynamical nature of the diffusion is a crucial process for stabilizing the neural network learning dynamics.

\textit{Experiment 3}:  In the next experiment we demonstrate that the neural network can learn and memorize while maintaining $\lambda$ close to the stable value of $1$. To do this we divide the neurons into two equally sized groups, $G_1$ and $G_2$, consisting of $500$ neurons each. This results in four groups of synapses: synapses that connect neurons within $G_1$, synapses from $G_1$ to $G_2$, synapses from $G_2$ to $G_1$ and synapses that connect neurons within $G_2$. 

We run the dynamics for a total of $160000$ time steps such that we have three distinct phases: pre-learning ($1\leq t \leq 80000$), learning ($80000 < t \leq 100000$) and post-learning ($t>100000$). In the pre-learning phase, the dynamics are as described in the previous section. The total resource $\mathcal{R}$ reaches a steady-state value and the eigenvalue $\lambda$ fluctuates near $1$ (see Figs.~\ref{fig:learning_lambda}, \ref{fig:learning_R}). 
In the learning phase, for neurons in group $G_{\nu}$ ($\nu=1 \mbox{ or } 2$) we modify \eqref{eq:sn_t_1} by introducing a time-dependent external stimulus, $\zeta_t^{(\nu)}$, 
\begin{align} \label{eq:sn_t_1_learning}
s_n^{t+1} = \twopartdef{1}{with prob.}{\sigma \left(\displaystyle\sum_{m=1}^{N} W_{nm}^t s_m^t + \zeta_t^{(\nu)} \right)} {0} {otherwise} {}{,}{.}
\end{align}
where $\nu$ is the group to which neuron $n$ belongs, and, letting $\zeta=0.15$, the learning protocol defining $\zeta_t^{(\nu)}$ is as shown in Fig.~\ref{fig:learning_protocol}. That is, starting at the beginning of the learning phase ($t=T_1=80000$), we stimulate neurons only in $G_1$; then, in the next time step, we stimulate neurons only in $G_2$; then, in the next two time steps, no stimulus is applied to either group; and this four step sequence is successively repeated until the end of the learning phase ($t=T_2=100000$), past which no stimuli are applied. As expected, sequential firing of $G_1$ neurons followed by $G_2$ neurons results in strengthening of excitatory synapses from $G_1$ to $G_2$ and weakening of excitatory synapses from $G_2$ to $G_1$. We plot the mean synaptic strength for the four groups of synapses in Figs.~\ref{fig:learning_excitatory}. Importantly, these learning-related changes in strengths of the four groups of synapses are preserved in the post-learning phase (after time step $100000$), thus confirming that the neural network remembered what it learned. 

Finally, Fig.~\ref{fig:learning_R} shows that during the learning phase there is a corresponding decrease in total resource $\mathcal{R}$. The increased resource consumption and the consequent decrease in $\mathcal{R}$ can be attributed to the increase in neuronal firing rates owing to the external stimulus. As the stimulus is removed in the post-learning phase, the plots in Figs. \ref{fig:learning_lambda}, \ref{fig:learning_R} show that the resource $\mathcal{R}$ is replenished and $\lambda$ resets to $1$ with fluctuations. Hence, in the post-learning phase we have balanced cortical state, and the neural network remembers what it learned. Thus, although the glial transport stabilizes a unique attracting macrostate with $\lambda \cong 1$, it, nevertheless, still potentially allows for distinct microstates representing stored information.

\begin{figure}
\centering{
\includegraphics[scale=0.31]{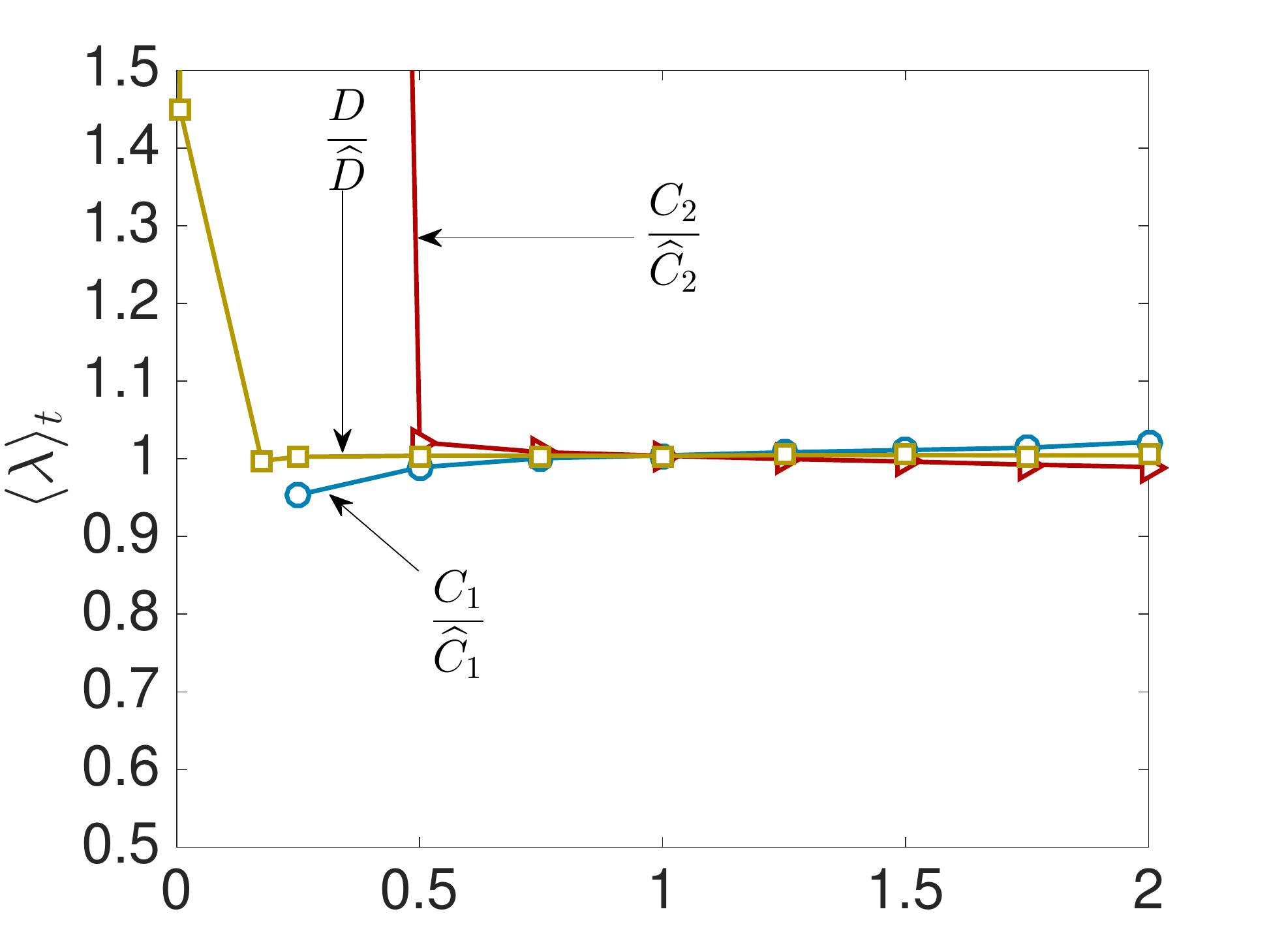}
}
\caption{\textit{Robustness of dynamics to parameter change}: Time average of the largest eigenvalue, $\langle \lambda \rangle_t$, as a function of $C_1/\widehat{C}_1$, $C_2/\widehat{C}_2$ and $D/\widehat{D}$ where $\widehat{C}_1$, $\widehat{C}_2$ and $\widehat{D}$ are the parameter values used for Fig.\ref{fig:stability_exp}-\ref{fig:learning}. All three curves show that our model is fairly robust to parameter changes, e.g., a $25$\% change in $C_1$ or $C_2$ yields a change in $\langle \lambda \rangle_t$ of about $0.3$\%}
\label{fig:robustness}
\end{figure}

\textit{Robustness}: We find that the qualitative results we obtain in our numerical experiments are fairly robust to parameter variations over a $25$\% range in $C_1$ and $C_2$ and even larger ranges for $D$ and $\bar{w}$. One indication of this is shown in Fig.~\ref{fig:robustness} where we plot the time averaged largest eigenvalue $\langle \lambda \rangle_t$ of $W$ as each parameter $C_1$, $C_2$ and $D$ normalized to their values used in Figs. \ref{fig:stability_exp}-\ref{fig:learning} is changed, while keeping the others fixed. We note that $\langle \lambda \rangle_t$ changes only by roughly $0.3$\% when $C_1$ or $C_2$ changes by 25\%.

\section{Conclusions}

A significant amount of the human body's energy consumption occurs at synapses in the brain \cite{kety:1957}. This energy is consumed by the biophysical mechanisms underlying transsynaptic signaling and in learning-related long-term changes in synapses \cite{suzuki:et:al:2011, newman:korol:gold:2011}. Metabolic resources maintaining this high rate of energy consumption are delivered by a network of glial cells, which transport these resources from the bloodstream to the synapses. Since these resources are key to the functioning of synapses, it is natural to ask if the structure and function of this resource transport network plays a role in controlling the activity of the neuronal network in a beneficial way \cite{nadkarni:jung:levine:2008, wade:et:al:2011, han:et:al:2013}. In this paper we have shown, using a two-layered network model comprised of glial cells and neurons, that the dynamics of metabolic resource transport across the network of glial cells can stabilize learning dynamics in neuronal networks. Specifically, our three numerical experiments showed that (i) the balance between supply and consumption of resources naturally leads to a regime in which excitation and inhibition are balanced, even if in the initial state they are not, (ii) in the absence of diffusion of resources across the glial cell network, the balanced state becomes unstable, and (iii) the glial regulated neuronal network can learn and subsequently return to the balanced state retaining the learned pattern. Furthermore, these findings are robust to parameter changes. 

It is well known that without homeostatic regulatory mechanisms that control synaptic strengths, Hebbian plasticity rules lead to runaway growth in synaptic efficacy and to excessive neural activity \cite{abbott:nelson:2000, miller:mackay:1994, watt:desai:2010}. Various regulatory mechanisms preventing this instability have been proposed theoretically \cite{miller:mackay:1994, morrison:aertsen:diesmann:2007, bienenstock:cooper:munro:1982, delattre:et:al:2015, zheng:dimitrakakis:triesch:2013, uhlig:et:al:2013} and found experimentally \cite{turrigiano:et:al:1998:watt:et:al:2000, ibata:sun:turrigiano:2008, markram:tsodyks:1996}. Synaptic scaling \cite{turrigiano:et:al:1998:watt:et:al:2000, turrigiano:2008} operates on a timescale of hours to days \cite{ibata:sun:turrigiano:2008} by reducing all the afferent synapses to a given neuron by the same amount so that relative differences in synaptic strengths are preserved. Mechanisms collectively known as homeostatic intrinsic plasticity modify the intrinsic excitability properties of a neuron depending on its firing activity \cite{zhang:linden:2003} by up- or down-regulating the expression of membrane ion channels. Another class of models considers plasticity rules whose parameters depend on neuronal activity (metaplasticity) \cite{bienenstock:cooper:munro:1982}. In addition, stand-alone STDP models that result in stable critical dynamics have been proposed \cite{morrison:aertsen:diesmann:2007, markram:tsodyks:1996}. The belief has been expressed that these diverse mechanisms operate together, complementing each other, to keep neural dynamics in a balanced regime (e.g., \cite{watt:desai:2010}). However, it has been pointed out recently that, consistent with our experiment 3, a homeostatic mechanism that operates on faster timescales might also be necessary \cite{zenke:hennequin:gerstner:2013}. In this paper, we hypothesize that the regulatory mechanism imposed by metabolic resource distribution via the glial network could play this role. For related work on the influence of the glial network on neurons see, e.g., Refs. \cite{magistretti:2006, vernadakis:1996, allen:barres:2005}. We note that regulation of network activity by depletion of metabolic resources was proposed recently by Delattre et al. \cite{delattre:et:al:2015}. Ref. \cite{delattre:et:al:2015} considers a modification of the STDP rule that depends on a global metabolic resource that is depleted by the globally averaged network activity, and finds that it results in stable neural activity. In contrast to globally-based resource regulation, we consider here a spatially distributed resource transportation network in which resource diffusion through the glial cell network plays the key role in stabilizing the neural dynamics at the balanced state ($\lambda$ near $1$ in our model). In the limit of infinitely fast diffusion, the resource at every glial cell would become the same, and our model would reduce to one similar to that of Ref. \cite{delattre:et:al:2015}. However, local resource depletion and transport through the network of glia is more realistic and allows one to study the effect of different forms of spatially structured neuron-glia interactions. Finally we note that while in this paper we used simple models for the neuronal network, the glial network, and the resource dynamics, our work can be extended to include more realistic modeling. 

In the broader context of network science, there has been much recent interest in multilayered networks \cite{kivela:2014, boccaletti:et:al:2014, domenico:et:al:2013} and in the dynamics of interdependent networks (e.g., power grid networks and the internet depend on each other \cite{buldyrev:et:al:2010}). Our work is an example of how the interactions of two different networks can result in beneficial dynamics, in particular, the feedback control stabilization of an otherwise disabling instability. Other examples of such interactions include coupled oscillator networks (a resource transport network regulating synchronization was considered in Ref.~\cite{skardal:et:al:2014}) and transportation networks \cite{parshani:et:al:2011}.

\end{document}